\documentclass[letter,twocolumn]{jpsj2} %% two-column layout

\newcommand{\msr}{$\mu$SR}
\newcommand{\cdos}{Cd$_2$Os$_2$O$_7$}
\newcommand{\cdre}{Cd$_2$Re$_2$O$_7$}

\title{Anomalous magnetic phase in an undistorted
pyrochlore oxide Cd$_2$Os$_2$O$_7$ induced by geometrical frustration}

\author{%
Akihiro \textsc{Koda}$^{1, 2}$\thanks{E-mail address: coda@post.kek.jp}, %
Ryosuke \textsc{Kadono}$^{1, 2}$, %
Kazuki \textsc{Ohishi}$^1$\thanks{Present address: Advanced Science Research Center, Japan Atomic Energy Agency}, 
Shanta R.~\textsc{Saha}$^1$, %
Wataru \textsc{Higemoto}$^3$, %
Shigeki \textsc{Yonezawa}$^4$, Yuji \textsc{Muraoka}$^4$\thanks{Present
address: Research Laboratory for Surface Science, Okayama University} and %
Zenji \textsc{Hiroi}$^4$}

\inst{$^{1}$Muon Science Laboratory, Institute of Materials Structure
Science, High Energy Accelerator Research Organization (KEK), Tsukuba, Ibaraki 305-0801\\
$^{2}$Department of  Materials Structure Science, The Graduate University
for Advanced Studies (SOKENDAI), Tsukuba, Ibaraki 305-0801\\
$^{3}$Advanced Science Research Center, Japan Atomic Energy Agency,
Tokai, Ibaraki 319-1195\\
$^{4}$Institute for Solid State Physics, University of Tokyo, Kashiwa, Chiba 277-8581}

\abst{%
We report on the muon spin rotation/relaxation study of a pyrochlore oxide, 
\cdos, which exhibits a metal-insulator (MI) transition at $T_{\rm MI}\simeq225$~K
without structural phase transition.
It reveals strong spin fluctuation ($\geq$10$^8$~s$^{-1}$)
below the MI transition, suggesting a predominant role of geometrical spin frustration
amongst Os$^{5+}$ ions.
Meanwhile, upon further cooling, an incommensurate spin density wave 
discontinuously develops below $T_{\rm SDW}\simeq150$~K.
These observations strongly suggest the occurrence of an anomalous magnetic transition 
and associated change in the local spin dynamics in undistorted pyrochlore 
antiferromagnet.}

\kword{$\mu$SR, strongly correlated electron, metal-insulator
transition, SDW, frustrated magnetism}

\begin{document}
\maketitle

Pyrochlore lattice is a three dimensional network of corner
shared tetrahedra, in which the corners are occupied by metallic ions.
It is well known as a stage of the geometrically
frustrated magnetism for the magnetic ions interacting antiferromagnetically.
A class of pyrochlore compounds, especially having 5$d$ transition metals, is
interesting as a testing ground for the implicit relation between charge
transfer and spin frustration, since 5$d$ electrons generally occupy
one of outer orbits far from the cation, and thus they have itinerant character.

A pyrochlore oxide, \cdos , has been known since 1974, as it is reported to 
exhibit metal-insulator (MI) transition at
$T_{\rm MI}\simeq225$~K~\cite{Sleight74,Mandrus01}.
The resistivity enhances approximately by 10$^3$ times upon cooling
below $T_{\rm MI}$.
A sharp step-like reduction of magnetic susceptibility at the MI
transition suggests occurrence of antiferromagnetic (AF) order below $T_{\rm MI}$.
Recent discovery of superconductivity in a related pyrochlore oxide, \cdre, has
triggered renewed attention to this compound.\cite{Hanawa01,Sakai01,Jin01}
From the viewpoint of electronic band structure, \cdos\ (5$d^3$)
differs from \cdre\ (5$d^2$) by one electron per transition metal.
In this regard, it is interesting to note that the recently discovered $\beta$-pyrochlore
osmate superconductors, AOs$_2$O$_6$ (A = K, Rb and Cs)\cite{Yonezawa04a, Yonezawa04b,Yonezawa04c}, are in a mixed valence state (5$d^{2.5}$).
While band structure calculations predict a
semi-metallic character for both \cdos\ and \cdre\cite{Singh02, Harima02},
\cdre\ exhibits successive structural transitions at $\sim$200~K
and $\sim$50~K with reducing temperature 
% leading to the occurrence of superconductivity, 
that may give rise to a drastic change on  both
electronic state and electron-phonon coupling.~\cite{Hiroi03}

Notably in the case of \cdos, no evidence is found for the structural change 
across the MI transition, according to X-ray diffractometry study on 
single crystals.\cite{Sleight74, Mandrus01}
More interestingly, while a slight change of the lattice constant (without changing
crystal symmetry) was
observed by the recent powder neutron diffraction study,
%while the crystal symmetry was unchanged.
no sign of magnetic Bragg peak was found at the lowest temperature of 
12~K~\cite{Reading01}.  A very recent attempt to clarify the magnetic
ground state by Cd-NMR was unsuccessful, as it turned out that the NMR signal 
disappeared below $T_{\rm MI}$.\cite{HArai}
This is in marked contrast to the case of the other compounds having
pyrochlore lattice, e.g., Fe$_3$O$_4$, where occurrence
of a static magnetic order at lower temperatures is well established,\cite{Verway39}
and thus motivated us to carry out muon spin rotation/relaxation (\msr )
measurements to clarify the magnetic ground state of \cdos .
In this Letter, we show that the ground state is an incommensurate 
spin density wave (SDW),
which appears at a temperature $T_{\rm SDW}\simeq$150~K 
considerably lower than $T_{\rm MI}\simeq225$~K.
Moreover, strong evidence for spin fluctuation probably due to the
geometrical frustration of local magnetic moments is found over the temperature 
region between $T_{\rm MI}$ and $T_{\rm SDW}$, where no sign of 
long-range order is observed.
The present result demonstrates the role of geometrical frustration competing
with the AF order state on the pyrochlore lattice.

\msr\ experiment was performed using a pulsed muon beam provided by 
the Muon Science Laboratory of KEK, Japan.  
Additional measurements were carried out at TRIUMF, Canada,
to obtain data with a time resolution better than the muon pulse width
at KEK ($\simeq50$~ns).
In both cases, a 100\% polarized muon beam with a momentum of 29~MeV/c was 
implanted into the specimen. 
Conventional \msr\ technique was employed with a He-flow type cryostat to
control the temperature of specimen down to $\sim$2~K.
For the measurements under a zero external field (ZF) condition, the residual field was compensated 
to be smaller than $10^{-5}$ T using a triple-axis Helmholtz coil system.
Subsequently, those under a longitudinal field (LF, up to 0.3~T applied along
the direction of the initial muon polarization) were performed to 
examine the influence of spin fluctuation. 
The time evolution of muon polarization was monitored by
measuring the decay positrons which were preferably emitted toward the 
muon spin direction upon their decay,  where the time-dependent asymmetry 
of positron events between a pair of forward/backward 
counters was proportional to the instantaneous muon polarization [$G_z(t)$].

\begin{figure}[tb]
\begin{center}
\includegraphics[width=1.0\linewidth]{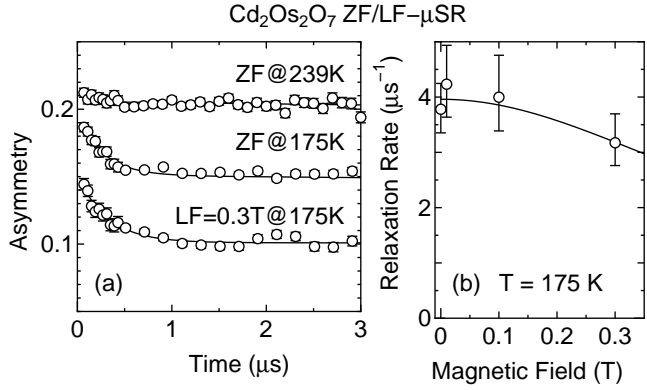}
\end{center}
\caption{(a) Time-dependent $\mu$-$e$ decay asymmetry ($\propto$
muon polarization) obtained at 239~K and
 175~K under zero external field, together with the result at 175~K
 under a longitudinal field of 0.3~T (KEK data).  Each spectrum is shifted (by 0.05) for 
 clarity,  where the full polarization corresponds to $\sim0.15$.
(b) Magnetic field dependence of muon spin relaxation rate at 175~K.
The solid curve represents the result of fitting analysis by the Redfield model.
}
\label{spectra_ht}
\end{figure}

Fig.~\ref{spectra_ht}(a) shows some examples of ZF-\msr\ time spectra observed 
at temperatures above and below $T_{\rm MI}\simeq 225$~K, together with 
that observed under a longitudinal field of 0.3~T below $T_{\rm MI}$.
As is found for the spectrum at 239~K, the implanted muons retains their initial polarization
between $T_{\rm MI}$ and ambient temperature.
%[The initial asymmetry ($A_0$) is $\sim0.14$, which is relatively small due to the
%experimental conditions.] 
However, upon lowering temperature below $T_{\rm MI}$, a slow muon spin relaxation is observed.
The field dependence of the muon spin relaxation rate ($\Lambda$) measured
at 175~K, which is determined by an analysis using a simple exponential
function,  
\begin{equation}
A_0G_z(t)=A\exp (-\Lambda t) + const.
\end{equation}
 is shown in Fig.~\ref{spectra_ht}(b), where the last term ($\simeq0.1$ and independent of time
 and field) represents a background signal from muons stopped in the sample holder:
 it was confirmed by the TRIUMF data that the entire fraction of 
 muons stopped in the sample exhibits depolarization by a rate $\Lambda$.
[The initial asymmetry ($A_0$) is relatively smaller than the usual
magnitude ($\simeq 0.2$) probably due to the formation of muonium
state and subsequent loss of polarization at the epithermal stage of muon implantation.]
It is established that the Redfield model is a good approximation
for evaluating $\Lambda$ under a longitudinal field, where we have
\begin{equation}
\Lambda (B) = \frac{2 \delta ^2 \nu}{(\gamma _{\mu} B)^2 + \nu ^2},
\label{redfield}
\end{equation}
with $\gamma _{\mu}$(=~$2\pi\times$~135.5~$\mu$s$^{-1}$/T) being the
gyromagnetic ratio of muon, $\delta$ the dipolar width, $\nu$ the
fluctuation rate of the local field, and $B$ the external field, respectively.
It is clear in Fig.~\ref{spectra_ht}(a) that the muon polarization does not recover even 
under a field of 0.3~T. This behavior strongly suggests the
fluctuation of local fields [namely, $\nu\gg\gamma _{\mu} B$ in eq.~(\ref{redfield})].
A fitting of the data in Fig.~\ref{spectra_ht}(b) using eq.~(\ref{redfield}) 
yields $\delta$ = 32(5)~$\mu$s$^{-1}$ [$\delta/\gamma _\mu\simeq$ 38(6)~mT] and $\nu$ =
$0.52(16)\times10^{9}$~s$^{-1}$, where these values should be regarded as lower
bounds considering the uncertainty of the fit.
This provides a natural explanation for the disappearance of Cd-NMR signal below
$T_{\rm MI}$, because the Cd nuclear moments would be subject to strong damping by the 
spin fluctuation.
Thus, the spin correlation just below $T_{\rm MI}$
is  characterized by the residual spin fluctuation at a rate of (or greater than) 
$10^8$--$10^9$~s$^{-1}$, 
which is in a stark contrast to the simple AF order currently presumed for \cdos.
It is plausible that the presence of such spin fluctuation is related
with the geometrically frustrated spin configuration (see below).

\begin{figure}[tb]
\begin{center}
\includegraphics[width=0.9\linewidth]{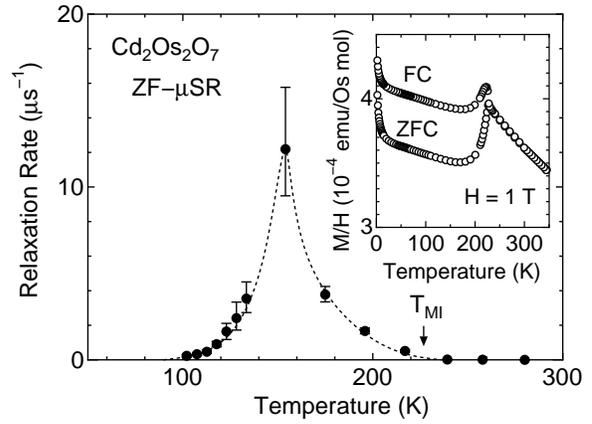}
\end{center}
\caption{Temperature dependence of muon spin relaxation rate under zero
 external field, where the data below $\sim$150 K correspond 
 $\lambda_2$ in eq.~(\ref{gzt}) (KEK data). The dashed curve is a guide to the eye.
Inset shows the result of dc magnetization measurements, which was obtained
for the specimen common to the \msr\ measurement.
}
\label{relax_ht-inset}
\end{figure}

\begin{figure}[tb]
\begin{center}
\includegraphics[width=1.0\linewidth]{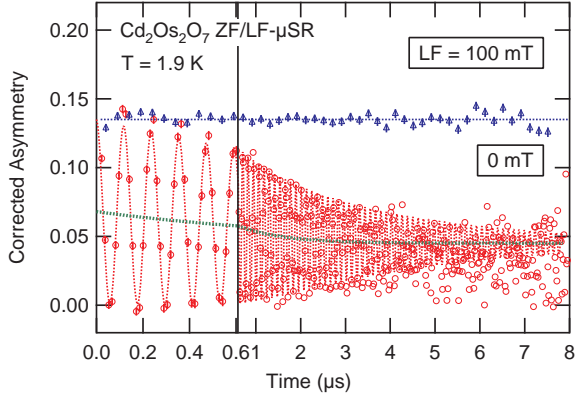}
\end{center}
\caption{\msr\ Time spectra at $T=1.9$~K under zero field (circles) and
 under a longitudinal field of 0.1~T (triangles), where the initial part of
 the spectra ($0\le t\le0.6$ $\mu$s) is magnified for clarity (TRIUMF data). 
 Dotted curves represent the result of fitting analysis.  A non-oscillating 
 component in the zero field spectrum deduced by the fitting is also 
 shown by thick dotts (see text).}
\label{2k-spectra}
\end{figure}

As shown in Fig.~\ref{relax_ht-inset}, $\Lambda$ exhibits a remarkable peak at
$\sim$150~K with a noticeable lambda-shaped tail typically seen for the critical slowing
down of spin fluctuation.
With further decreasing temperature below $\sim$150~K, a spontaneous muon spin
precession signal is observed (see Fig.~\ref{2k-spectra}).
These observations apparently suggest a drastic change of the magnetic
correlation time at
%%the occurrence of  a quasi-static magnetic order at
$\sim$150~K.  It is interesting to compare the temperature dependence of 
$\Lambda$ with that of dc susceptibility shown in the inset of
Fig.~\ref{relax_ht-inset} obtained on the same specimen used
for the \msr\ measurement.
The susceptibility exhibits a pronounced peak at $T_{\rm MI}$ as previously
reported,\cite{Sleight74, Mandrus01} whereas no clear anomaly is
observed at $\sim$150~K where a magnetic transition is inferred
from the present \msr\ study.
A similar situation is suggested for the specific heat, where the anomaly is 
observed only at $T_{\rm MI}$.~\cite{Mandrus01}
%%Considering that $T_{\rm MI}$ corresponds to the temperature where 
%%the increase of $\Lambda$ sets in,
%%the result suggests that 
%%the occurrence of the truly static magnetic order is
%%shifted down from $T_{\rm MI}$ to $\sim$150 K by geometrical frustration.
Thus, while it bears some features common to conventional magnetic
order, microscopic detail of the transitions observed by \msr\ seems to be 
quite unconventional.

Now we focus on the nature of magnetically ordered state below 150 K.
It is clear in Fig.~\ref{2k-spectra} that a spontaneous precession of muon polarization 
under zero external field is observed at 1.9~K. This unambiguously demonstrates the 
development of a static internal field associated with long-range magnetic 
order.
In the case of uniform magnetic ordering in a powder specimen, 1/3 of
the implanted muons shows no relaxing behavior, since the direction of
the internal field coincides with its polarization.
However, the ZF-\msr\ spectra in Fig.~\ref{2k-spectra} has a
character that the center of the precession signal 
exhibits an exponential damping with an initial value greater than 1/3 of
the total asymmetry. [They were obtained at TRIUMF, where the use of muon 
{\sl veto} counter as a sample holder allowed us to obtain background-free spectra (i.e., 
$const.$ in eq.(1) is zero).] 
%% expected for powder specimen with uniform magnetic order.
This suggests that the local field have a strong spatial modulation, where 
the internal field is much reduced from the maximal value over a certain fraction
of the total volume due to incommensurate distribution of local moment size. 
The absence of depolarization in the \msr\ time spectrum under a longitudinal
field of 0.1~T (see Fig.~\ref{2k-spectra}) eliminates the possibility that 
the exponential damping is that of the longitudinal component due to residual spin fluctuation.
Another possibility of macroscopic phase separation into AF and non-magnetic
phases is also ruled out by the fact that the previous powder neutron diffraction 
study was unsuccessful to identify the magnetic order down to 12~K.\cite{Reading01} 
Thus, we are led to conclude that the magnetic ground state of \cdos\ is a quasi-static
and incommensurate spin density wave (SDW) below $T_{\rm SDW}\simeq150$~K.

\begin{figure}[tb]
\begin{center}
\includegraphics[width=1.0\linewidth]{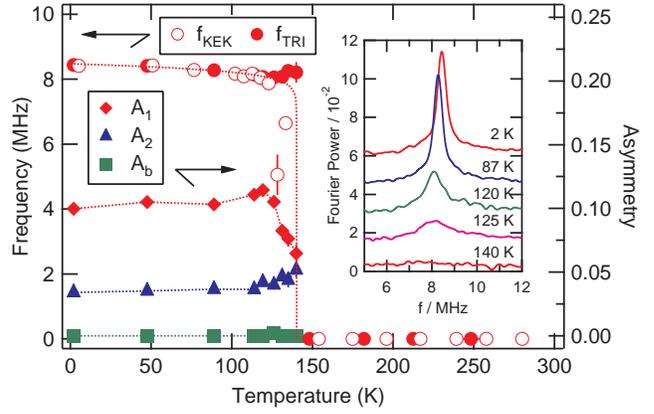}
\end{center}
\caption{Temperature dependence of the muon spin precession
 frequency, where open and solid circles represent data 
on the same specimen obtained at KEK ($f_{\rm KEK}$) and TRIUMF ($f_{\rm TRI}$), respectively. 
The onset temperature of SDW inferred from $f_{\rm KEK}$ has more 
uncertainty due to the limited time resolution.
The amplitudes of oscillating ($A_1$), non-oscillating
 ($A_2$), and the background ($A_{\rm b}$) components are displayed for the TRIUMF data.
(Dashed curves are guides to the eye.)
Inset: Fourier amplitude of ZF-\msr\ spectra at typical temperatures.} 
\label{frq_at_lt}
\end{figure}

For the quantitative discussion, the ZF-\msr\ time spectra below $T_{\rm SDW}$ 
(obtained at TRIUMF) are
analyzed by using the following formula for powder specimen,
\begin{eqnarray}
A_0 G_z (t) 
%%= \sum_{i=1,2}A_i \left[\frac{1}{3}+\frac{2}{3}\exp(-\lambda _i t
%%		   )\cos(2\pi f_i t + \phi_i)\right]+ A_{\rm b},
&= &A_1 \left[\frac{1}{3}+\frac{2}{3}\exp(-\lambda _1 t )\cos(2\pi f t +
       \phi)\right] \nonumber\\
 & &      + A_2 \left[\frac{1}{3}+\frac{2}{3}\exp(-\lambda _2
       t)\right] + A_{\rm b},
\label{gzt}
\end{eqnarray}
%\begin{equation}
%A_0 G_z (t) 
%= A_1 \left[\frac{1}{3}+\frac{2}{3}\exp(-\lambda _1 t )\cos(2\pi f t +
%       \phi)\right] + A_2 \left[\frac{1}{3}+\frac{2}{3}\exp(-\lambda _2
%       t)\right] + A_{\rm b},
%\label{gzt}
%\end{equation}
where the total asymmetry $A_0$ 
%($\sim0.14$) is the initial asymmetry which 
is split into two signals with $A_i$ being thier partial asymmetry ($i=1,2$)
and a time-independent background $A_{\rm b}$, $\lambda _i$ is the relaxation
rate, and $f=\gamma_\mu B/2\pi$ is the precession frequency with 
$B$ being the spontaneous local field.
Eq.~(\ref{gzt}) corresponds to the model in which the density distribution of the
internal field [$n(B)\propto\int G_z(t)\exp(-i\gamma_\mu Bt)dt$] is
approximated by a square wave pattern with the duty ratio being $A_1/(A_1+A_2)$.
The temperature dependence of $A_i$, $A_{\rm b}$, and $f$ is shown in
Fig.~\ref{frq_at_lt}.
Unlike the case of conventional magnetic phase transition, 
%the internal field ($\propto f$) exhibits a discontinuous change at $T_{\rm SDW}$.
the development of the internal field is quite steep, as $f$ exhibits 
a discontinuous change at $T_{\rm SDW}$ with least dependence 
on temperature immediately below $T_{\rm SDW}$.
It is thus suggested that the transition may not be driven solely by 
magnetic interaction but some other degrees of freedom.
The relative yield of two components is mostly independent of temperature,
where $A_1/(A_1+A_2)\simeq3/4$.
Based on the two component approximation, we obtain that the volume-averaged
internal field ($2\pi f A_1/\gamma_\mu\sum A_i$) is $\sim$46~mT at 1.9~K, which
is fairly close to the dipolar width ($\delta/\gamma_\mu$) deduced at 175~K.
%It is suggested that the resulted static SDW ground state has an
%intimate relation to the dynamical nature seen just below $T_{\rm MI}$
%presumably due to the spin frustration.

It would be worth mentioned at this stage that a slight but distinctive change in the temperature
dependence of various parameters has been indeed observed at $T_{\rm SDW}$ in the
earlier neutron diffractometry~\cite{Reading01}.
Furthermore, the temperature dependence of electric resistivity
shows a hump (or a narrow plateau) at $T_{\rm SDW}$ regardless of crystal quality\cite{Sleight74, Mandrus01}.
%Even a narrow plateau was observed in the case of a
%polycrystalline powder sample.
The activation energy estimated from the temperature
dependence of resistivity starts to drop to zero with decreasing
temperature below $T_{\rm SDW}$\cite{Mandrus01}.
These anomalies coherently point to the presence of a transition 
that is revealed by \msr\ experiment.

%We would like to point out that the absence of anomaly
%concerning muon spin relaxation rate at $T_{\rm MI}$ does not eliminate the
%possibility of antiferromagnetic ordering.
%It was confirmed by our \msr\ experiments under a high transverse field of
%1~T that the hyperfine coupling constant due to the uniform
%magnetization is negligibly small.
%In general, if a muon stopped at an interstitial site which is symmetric
%to Os-sites, the dipolar field is partially canceled.
%Even under such a spin configuration, muons sensitively feel spin
%luctuation.
Here, we discuss the local configuration of Os moments relative
to muons, and possible models to interpret the results of \msr\ and
other measurements including magnetic susceptibility in a coherent manner.
Provided that muon occupies the center of an Os tetrahedron 
to form an inner ${\mu}^{+}$O$^{2-}_6$ octahedron suitable for accommodating a positively
charged muon, which being also consistent with small muon depolarization rate 
at ambient temperature (see Fig.~\ref{spectra_ht}a) as it is estimated to be $2.5\times10^{-3}$ MHz 
from random local fields of Cd nuclear magnetic moments, 
the hyperfine coupling constant is
calculated to be $\delta/\gamma_\mu\simeq0.21$~T/${\mu}_B$ (with ${\mu}_B$
being the Bohr magneton). 
The comparison of $\delta$ with that deduced at 175~K 
leads to the estimated average moment size of 0.18${\mu}_B$ per Os
ion which is much smaller than that expected for completely localized
5$d$ electrons.  This is again in line with the presence of strong magnetic
frustration of electrons that often leads to the shrinkage of effective local moment.
In such a situation, it is commonly observed in many spinel and
pyrochlore oxides that structural transition to a lower crystal symmetry occurs to lift the
geometrical degeneracy.  Interestingly, however, no indication of structural transition 
is reported for the case of \cdos.\cite{Reading01} 
This may have a relevance to the fact that the $t_{2g}$ manifold in the
electronic state is fully occupied by 5$d$ electrons (5$d^3$) so that the Jahn-Teller
like distortion may not be favored.  In any case, the geometrical frustration and associated
fluctuation of local spins naturally explains the absence of truly
static magnetic order for $T_{\rm SDW}\le T\le T_{\rm MI}$, as it is
suggested by \msr .

On the other hand, the susceptibility clearly indicates the
development of quasi-static magnetization below $T_{\rm MI}$.
Provided that the Os ions have an Ising-like single ion anisotropy, 
this seemingly contradictory behavior is understood by considering
a complex magnetic order parameter consisting of static and dynamically fluctuating 
components.  While the absence of orbital degeneracy disfavors such an 
anisotropy, it might be caused by a higher order exchange interaction.
Then, one possibility is that the internal fields due to the quasi-static component
(e.g., along [111] axis) may cancel with each other at the muon site due to the
symmetric spin configuration, while the fluctuating (transverse) component
leads to complete muon depolarization.

Another intriguing possibility would be a situation similar to
the partially disordered 
(PD) phase observed in the Ising antiferromagnets on the two-dimensional 
triangular lattice (2D-TL), 
where one of the sublattices remains disordered due to the second
nearest exchange interaction and dynamically replaces with each 
other.\cite{Mekata:77,Kaburagi:81,Takayama:83,Fujiki:84,Ueno:85}
Very recently, an interesting observation similar to the present case 
has been reported on NaCrO$_2$ (a candidate for 2D-TL consisting of the
edge-sharing CrO$_6$ octahedra), where a quasi-static 
internal field detected by \msr\ and Na-NMR measurements develops 
only over a lower temperature range 
that is far below the temperature at which the
susceptibility and specific heat exhibit an anomalous peak.~\cite{Olariu06} 
The wipe-out of NMR signal at the intermediate temperature strongly suggests 
the presence of fast spin fluctuation, whereas its recovery at further lower 
temperatures is explained by the freezing of the fluctuating spins.
Despite the difference in the lattice dimensionality between those systems, the
observed tendency suggests a possible common background between 2D-TL and the
pyrochlore.  It might be also interesting to note that they share a feature of 
completely filled $t_{2g}$ orbitals ($3d^3$ for Cr$^{3+}$) that may keep the system 
away from the Jahn-Teller like distortion.  Unfortunately, this in turn give rise to 
the absence of anisotropy in the exchange interaction: 
such anisotropy is presumed to be a key prerequisite for the PD phase in 2D-TL.

In the case of NaCrO$_2$, they discuss possible role of so-called 
``$Z_2$ vortices" predicted for the 2D Heisenberg antiferromagnets,\cite{Kawamura:84}  
where the dynamical character of the $Z_2$ vortices leads to 
spin fluctuation over an intermediate temperature range, 
while a short-range order due to the pairing of $Z_2$ vortices
is anticipated at lower temperatures.  
The present result on \cdos\ seems to
share a remarkable similarity with the behavior predicted for the 2D-TL case.  
However, it is clear that the 
three-dimensional version of the theoretical model is needed
to understand the microscopic property of the Heisenberg antiferromagnets on the
pyrochlore lattice.

In summary, it is inferred from \msr\  measurements in \cdos\ that strong spin fluctuation persists
below the MI transition, which is followed by the transition to
a quasi-static SDW ground state with decreasing temperature below $T_{\rm SDW}\simeq150$~K.
These observations suggest the predominant role of geometrical frustration 
and occurrence of anomalous magnetic ground state on the undistorted pyrochlore
lattice.  Further study including the development of appropriate theoretical model for the
pyrochlore magnets is required for the quantitative understanding of the
ground state.

We would like to thank technical support by the staff of KEK and TRIUMF.
We acknowledge the helpful discussion with T.~Kamiyama, K.~Ishida and S. Takeshita.
This work was partially supported by a Grant-in-Aid for Scientific
Research on Priority Areas and a Grant-in-Aid for Creative Scientific
Research from the Ministry of Education, Culture, Sports, Science and
Technology, Japan.
One of the authors (A.K.) acknowledges support of JSPS Research Fellowships for Young
Scientists.

\end{document}